\documentstyle[aps,prl,epsf,twocolumn]{revtex}
\begin{document}
\draft

\wideabs{
\title{Residual correlation in two-proton interferometry 
from lambda-proton strong interactions}

\author{Fuqiang Wang}
\address{Nuclear Science division, Lawrence Berkeley National Laboratory,
Berkeley, CA 94720, USA}

\maketitle

\begin{abstract}
We investigate the residual effect of $\Lambda p$ strong interactions
in $pp$ correlations with one proton from $\Lambda$ decays.
It is found that the residual correlation is about 10\% of 
the $\Lambda p$ correlation strength, and has a broad distribution 
centered around $q \approx 40$~MeV/$c$.
The residual correlation cannot explain the observed structure 
on the tail of the recently measured $pp$ correlation function
in central Pb+Pb collisions by NA49 at the SPS.
\end{abstract}

}


It has been recently shown, by using an Urbana-type potential,
that the lambda-proton ($\Lambda p$) strong interaction induces 
a large peak in the correlation function~\cite{lambda_p}.
Because $\Lambda$ decays into $\pi^-p$ with low decay momentum
($Q \simeq 100$~MeV/$c$) and large branch ratio (64\%), 
the strong $\Lambda p$ correlation should have an effect on two-proton 
($pp$) correlations if one of protons comes from $\Lambda$ decays. 
In this brief report, we investigate the magnitude of such effect.

We shall derive analytically the residual correlation following 
straightforward kinematics.
Let $k$ denote the momentum magnitude of either particle 
in the $\Lambda p$ pair rest frame (Fig.~\ref{fig:decay}a).
$\Lambda$ decays into a proton (denoted as $p_{\Lambda}$) and 
a $\pi^-$, each with a momentum magnitude $Q$ in the $\Lambda$ 
rest frame (Fig.~\ref{fig:kine}b). 
Since both $Q$ and $k$ ($^<_{\sim} 50$~MeV/$c$, where the $\Lambda p$ 
correlation is significant) are small compared to 
proton or $\Lambda$ rest mass, 
for simplicity, we shall treat the problem non-relativistically.
Let $q$ be one half of the invariant relative momentum between $p$ and 
$p_{\Lambda}$, then $q^2 = a^2k^2-Qak\cos\theta+Q^2/4$, where $\theta$
is the $\Lambda$ decay angle, and $a=(1+m_p/m_{\Lambda})/2\approx 0.92$.
The kinematically allowed region of $q$ versus $k$ is sketched
in Fig.~\ref{fig:kine} as shaded area. For a fixed $q$, the allowed 
$k$ range is $k_- \leq k \leq k_+$, with $a^2k^2_{\pm}=(q \pm Q/2)^2$.

The following relation holds for pair multiplicity distributions:
\begin{equation}
\frac {d^2N} {dqdk} = \frac {2q} {Qak} \frac {d^2N} {dkd\cos\theta}
\propto \frac {qk} {Q} C_{\Lambda p}(k).
\end{equation}
Here $d^2N/dkd\cos\theta \propto k^2$ assuming isotropic decay
of unpolarized $\Lambda$, and $C_{\Lambda p}(k)$ is the $\Lambda p$
correlation function.
The $pp_{\Lambda}$ correlation function is therefore
\begin{equation}
C_{pp_{\Lambda}}(q) =
\frac {\int_{k_-}^{k_+} kC_{\Lambda p}(k) dk} {\int_{k_-}^{k_+} k dk}.
\end{equation}

Without an explicit form for $C_{\Lambda p}(k)$, one cannot go further.
Motivated by the $\Lambda p$ correlation results in Ref.~\cite{lambda_p},
we assume a Gaussian form for $C_{\Lambda p}$,
\begin{equation}
C_{\Lambda p} = 1 + \lambda e^{-k^2/2k_0^2}.
\label{eq:gauss}
\end{equation}
Thus,
\begin{equation}
C_{pp_{\Lambda}}(q) = 1 + \frac {\lambda a^2k_0^2} {qQ} \left[ 
e^{-\frac{(q-Q/2)^2}{2a^2k_0^2}} - e^{-\frac{(q+Q/2)^2}{2a^2k_0^2}} \right].
\label{eq:c2}
\end{equation}
Quick examination of Eq.~(\ref{eq:c2}) indicates that the resulting 
$pp_{\Lambda}$ residual correlation peaks at $q \approx Q/2$,
and its amplitude is reduced, due to phase-space population,
by a factor of $Q^2/2a^2k_0^2$ from the original $\Lambda p$ 
correlation strength.

NA49 experiment at the SPS has recently measured $pp$ correlation function 
at midrapidity in central Pb+Pb collisions at 158~AGeV~\cite{na49}. 
The proton pair sample (in the NA49 acceptance) 
was contaminated by weak decay protons resulting
in 44\% pairs containing at least one weak decay proton.
A Gaussian source size $R_g = 3.85 \pm 0.15$~(stat.)~fm was extracted.
A statistically significant bump structure was observed 
at $q \approx 70$~MeV/$c$ in the correlation function.

To estimate the magnitude of the residual correlation at SPS energy,
we calculate $\Lambda p$ correlation function for a $R_g = 3.85$~fm
source (for both $\Lambda$ and proton),
the results of which is shown in Fig.~\ref{fig:c2}a.
For computational reasons, we have used thermal momentum distributions 
of temperature $T=3$~MeV.
The difference in the correlation functions between $T=3$ and 300~MeV
(the observed inverse slope of proton transverse mass distributions) 
is less than 5\%~\cite{lambda_p}.
We fit the correlation function to Eq.~(\ref{eq:gauss})
yielding $\lambda=0.62$ and $k_0=21$~MeV/$c$.
The calculated residual correlation function of Eq.~(\ref{eq:c2}) 
is superimposed in Fig.~\ref{fig:c2}a. 
The residual correlation function peaks at $q \approx 40$~MeV/$c$ 
with amplitude 0.05 ({\it i.e.}, a factor of $\sim$10 reduction).

$\Sigma^+ p$ case can be treated in the same way as $\Lambda p$, 
except that the $\Sigma^+ p$ correlation has contributions from
the relative Coulomb interactions.
Further because of the large $\Sigma^+$ decay momentum (190~MeV/$c$), 
the resulting residual correlation in $pp_{\Sigma^+}$ is negligible.

Some of the $\Lambda$'s are from decays of $\Sigma^0$'s and
multistrange baryons.
Correlations between proton and these parent particles should 
also leave residual correlations between the proton and decay protons.
Take $\Sigma^0 p$ as an example. 
The allowed kinematic region of $q$ versus $k$ 
(of the $\Sigma^0 p$ system) does not have the sharp wedges at 
$(k,q)=(0,Q/2)$ and $(Q/2a,0)$ as shown in Fig.~\ref{fig:kine}.
Instead, they are smeared by $\pm Q_{\Sigma^0}/2$, where 
$Q_{\Sigma^0}\approx 74$~MeV/$c$ is the $\Sigma^0$ decay momentum. 
As a result, the magnitude of the residual $pp_{\Sigma^0}$ correlation 
is negligible compared to that of $pp_{\Lambda}$.
(The $\Lambda$ and $\Sigma^0$ production abundances should 
be comparable in high energy nucleus-nucleus collisions.)

Experimentally measured protons are a mixture of direct protons 
(produced + primordial) and weak decay protons. 
Due to the strong and Coulomb interactions, correlations between
direct protons display a dip (at $q=0$) 
and peak (at $q\approx 20$~MeV/$c$) structure~\cite{koonin}.
The $pp$ correlation function for a $R_g=3.85$~fm source for direct 
protons is shown as the thin solid curve in Fig.~\ref{fig:c2}b. 
Except the residual correlation inherited from the parent particles,
no further correlations exist between direct protons and weak decay protons.
To estimate the effect of the residual correlation on the measured
$pp$ correlation function, for simplicity, we assume that all 
background pairs contain one proton from $\Lambda$ decays.
The combined $pp$ correlation function, $0.56C_{pp}+0.44C_{pp_{\Lambda}}$, 
is shown as the thick solid curve in Fig.~\ref{fig:c2}b.
We conclude that the residual correlation cannot explain the observed
structure at $q \approx 70$~MeV/$c$ by NA49~\cite{na49}. 
It should be noted that NA49 has large errors in the momentum 
determination of weak decay protons~\cite{toy,na49nim}, therefore 
the residual correlation is significantly smeared (weakened).

In order to check the effect of the NA49 acceptance~\cite{na49}, 
we generate $\Lambda$'s and protons according to the 
experimentally measured transverse spectra~\cite{bormann,stopping}, 
and a flat rapidity distribution, and isotropically decay the 
$\Lambda$'s into protons with the proper lifetime and branch ratio.
The momenta of the $\Lambda$ decay protons are calculated 
from the kinematics, and no attempt is made to incorporate 
the reconstruction errors of NA49 for these momenta. 
The absolute normalization of the proton abundance is taken from 
Ref.~\cite{stopping}. The relative abundance of $\Lambda$ to proton
is such that 44\% of the resulting proton pairs in the acceptance 
are $pp_{\Lambda}$ pairs.
The source size is kept as $R_g = 3.85$~fm.
The $\Lambda p$ correlation is calculated using the protons in the 
acceptance and the $\Lambda$'s whose decay protons are in the acceptance. 
The result is consistent with the $C_{\Lambda p}$ shown in 
Fig.~\ref{fig:c2}a, which was obtained using a static thermal
source of $T=3$~MeV.
This agrees with the observation in Ref.~\cite{lambda_p}.
We repeat the calculation of $pp$ correlation using the
protons (including the $\Lambda$ decay protons) in the acceptance.
For a $pp_{\Lambda}$ pair, the correlation weight is $C_{\Lambda p}$ 
and is associated with the $q$ value of the $pp_{\Lambda}$ pair.
The $pp$ correlation result is consistent with the one
shown in Fig.~\ref{fig:c2}b (the thick curve).

As shown in Fig.~\ref{fig:c2}, the residual correlation fills in 
the dip at $q \approx 40$~MeV/$c$ of the $pp$ correlation function.
If there were no correlation between $\Lambda$ and proton 
($C_{\Lambda p}=1$), then the combined $pp$ correlation would be 
the dashed curve in Fig.~\ref{fig:c2}b.
The major effect of $\Lambda p$ correlation in the $pp$ correlation
function is seen in the $q \approx 40$~MeV/$c$ region.
It has been shown that this region is sensitive to the elongation 
and lifetime of the proton source~\cite{lifetime}.
The residual correlation may complicate the interpretation of such 
information extracted from $pp$ correlation functions at high energies, 
where $\Lambda$ production cannot be neglected.

In summary, we have investigated the residual effect of $\Lambda p$ strong 
interactions in $pp$ correlations with one proton from $\Lambda$ decays.
It is found that the residual correlation is significant,
on the order of 10\% of the $\Lambda p$ correlation strength, 
and has a broad distribution centered around $q \approx 40$~MeV/$c$.
We conclude that the structure observed at $q \approx 70$~MeV/$c$ 
in the $pp$ correlation function measured by NA49 cannot be explained 
by the residual correlation discussed here. 
We note that the residual correlation affects most the 
$q \approx 40$~MeV/$c$ region of $pp$ correlation function,
which may complicate the usage of the region to extract 
time information about the proton source.

\acknowledgements{
The author acknowledges G. Cooper, R. Lednicky, A.M. Poskanzer, 
M. Toy and N. Xu for fruitful discussions. 
He also thanks S.~Pratt for reading of the manuscript.
This work was supported by the U.S. Department of Energy 
under contract DE-AC03-76SF00098.



\newpage

\begin{figure}
\centerline{\epsfxsize=0.5\textwidth\epsfbox[60 120 520 400]{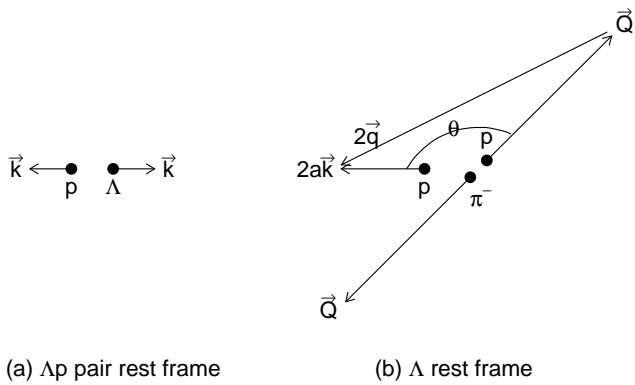}}
\caption{Non-relativistic view of a $\Lambda p$ pair in the pair rest 
frame (a), and a $\Lambda p$ pair and $\Lambda\rightarrow\pi^-p$ decay 
in the $\Lambda$ rest frame (b).}
\label{fig:decay}
\end{figure}

\begin{figure}
\centerline{\epsfxsize=0.4\textwidth\epsfbox[90 100 460 430]{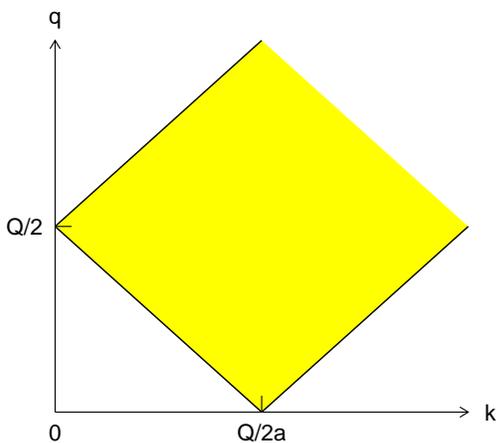}}
\caption{Allowed kinematic region in the non-relativistic approximation
for $q$ (one half of the relative momentum of the $pp_{\Lambda}$ pair)
versus $k$ (one half of the relative momentum of the $\Lambda p$ pair). 
$Q$ is the $\Lambda$ decay momentum. $a=(1+m_p/m_{\Lambda})/2$.}
\label{fig:kine}
\end{figure}

\begin{figure}
\centerline{\epsfxsize=0.5\textwidth\epsfbox[30 0 480 530]{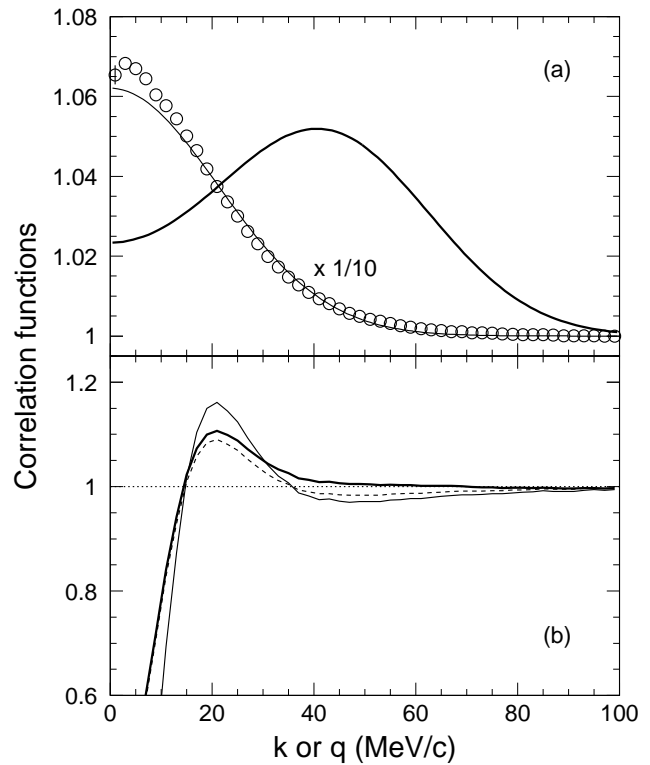}}
\caption{Upper panel: $\Lambda p$ correlation function ($C_{\Lambda p}$) 
for a $R_g=3.85$~fm source is shown in circles. 
The correlation strength, $C_{\Lambda p}-1$, has been divided by 10 
for clarity. The thin curve is a Gaussian fit to $C_{\Lambda p}$.
The thick curve is the resulting residual correlation 
($C_{pp_{\Lambda}}$) between the proton and the $\Lambda$ decay proton. 
Lower panel: The thin solid curve shows the $pp$ correlation function 
($C_{pp}$) for a $R_g=3.85$~fm source. The thick solid curve shows 
the $pp$ correlation function after including the residual correlation, 
$0.56C_{pp}+0.44C_{pp_{\Lambda}}$, with 44\% contamination from 
$\Lambda$ decay protons in the $pp$ pair sample.
The dashed curve shows the $pp$ correlation function if there were 
no correlation between $\Lambda$ and proton 
({\em i.e.,} $C_{\Lambda p}=1$ and $C_{pp_{\Lambda}}=1$).}
\label{fig:c2}
\end{figure}


\begin{references}

\bibitem{lambda_p}
F.~Wang and S.~Pratt, Phys. Rev. Lett., in print. nucl-th/9907019 (1999).

\bibitem{na49}
H. Appelsh\"{a}user {\it et al.} (NA49 Coll.), nucl-ex/9905001 (1999).

\bibitem{koonin}
S.E.~Koonin, Phys.~Lett. {\bf 70B}, 43 (1977).

\bibitem{toy}
M. Toy, PhD thesis, UCLA, 1999.

\bibitem{na49nim}
S.~Afanasiev {\it et~al.} (NA49~Coll.), 
Nucl. Instrum. Meth. {\bf A430}, 210 (1999).

\bibitem{bormann}
C.~Bormann {\it et~al.} (NA49 Coll.), J. Phys. G {\bf 23}, 1817 (1997).

\bibitem{stopping}
H.~Appelsh\"{a}user {\it et~al.} (NA49 Coll.), 
Phys. Rev. Lett. {\bf 82}, 2471 (1999).

\bibitem{lifetime}
S. Pratt and M.B Tsang, Phys. Rev. C {\bf 36}, 2390 (1987).

\end{references}
\end{document}